\documentclass[twocolumn,usenatbib,superscriptaddress,showpacs,prd,aps,amsmath,amssymb,nofootinbib]{mnras}
\usepackage{natbib}
\usepackage{graphicx,color}
\usepackage{amsmath,amssymb}
\usepackage{verbatim}
\usepackage{hyperref}
\usepackage{float}
\usepackage{wasysym}
\usepackage{amssymb,graphicx}
\usepackage{epsfig}
\usepackage{psfrag}
\usepackage{dsfont}
\usepackage{amsfonts}
\usepackage{mathrsfs}
\usepackage{multirow}
\usepackage{times}
\usepackage{soul}
\usepackage[shortlabels]{enumitem}

\newcommand{\vk}{v_{\rm kick}}
\newcommand{\Mbh}{M_{\bullet}}

\newcommand{\jzjc}{j_z/j_{\rm circ}}
\newcommand{\vc}{v_{\rm circ}}

\newcommand{\kmpersperpc}{km $\rm{s}^{-1} \rm{pc}^{-1}$}

\title[Eccentric Nuclear Disks Caused by GW Recoil Kicks]{Counter-Rotation and Slow Precession in Aligned Eccentric Nuclear Disks due to Gravitational Wave Recoil Kicks}

\author[J. C. Bright, T. Akiba, A.-M. Madigan]{Jane C. Bright$^{1,2}$ \thanks{Contact e-mail: janebright@arizona.edu}, Tatsuya Akiba$^{2}$, \& Ann-Marie Madigan$^{2}$%
\\
$^{1}$Department of Astronomy and Steward Observatory, University of Arizona, Tucson, AZ
\\
$^{2}$JILA and Department of Astrophysical and Planetary Sciences, University of Colorado, Boulder, CO}

\begin{document}
\maketitle


\begin{abstract}
The M31 nucleus contains a supermassive black hole embedded in a massive stellar disk of apsidally-aligned eccentric orbits. It has recently been shown that this disk is slowly precessing at a rate consistent with zero. Here we demonstrate using $N$-body methods that apsidally-aligned eccentric disks can form with a significant ($\sim$0.5) fraction of orbits counter-rotating as the result of a gravitational wave recoil kick of merging supermassive black holes. Higher amplitude kicks map to a larger retrograde fraction in the surrounding stellar population which in turns results in slow precession. We furthermore show that disks with significant counter-rotation are more stable (that is, apsidal-alignment is most pronounced and long lasting), more eccentric, and have the highest rates of stars entering the black hole's tidal disruption radius.  
\end{abstract}

\begin{keywords}
galaxies: kinematics and dynamics $<$ Galaxies, galaxies: nuclei $<$ Galaxies, stars: kinematics and dynamics $<$ Stars, transients: tidal disruption events $<$ Transients
\end{keywords}

\section{Introduction}

The nucleus of the Andromeda galaxy (M31) contains a double-peaked stellar light profile, well explained by an apsidally-aligned, eccentric stellar disk in orbit about a supermassive black hole \citep{Light:1974, Lauer:1993, Lauer:1998, Tremaine:1995, Statler1999, Kormendy1999, Bacon2001, Sambhus2002, Peiris:2003,Bender2005}. 
While differential apsidal precession between orbits of varying angular momentum and energy should destabilize such disks, inter-orbit gravitational torques \citep{Rauch1996} act to maintain the apsidal-alignment \citep{Madigan:2018}. 
One might expect the disk to precess at a rate $\sim (M_{\rm disk}/{\Mbh}) ~P^{-1}$, where $M_{\rm disk}$ is the mass of the disk, $\Mbh$ is the black hole mass, and $P$ is the orbital period of a typical star. For ${\Mbh}/M_{\rm disk} \approx \mathcal{O}{(10)}$ (corresponding to $\Mbh = 10^8 M_{\odot}$ as in M31) \citep{McConnachie:2005, Bender2005}, this yields a rate of $\sim\!65$ \kmpersperpc. 
A wide range of precession values, from 3 - 34 \kmpersperpc, have been found in simulations of the M31 nuclear disk using a variety of techniques \citep{Bacon2001, Jacobs2001, Salow2001, Sambhus2002, Salow2004}. 
However, \citet{Lockhart:2018} have shown that the lopsided disk in the Andromeda nucleus has a precession rate consistent with zero: $0 \pm 3.9$ \kmpersperpc. This rate is in agreement with the formation mechanism for the young circular disk of stars in the M31 nucleus as put forward in \cite{Chang:2007}, see also \citet{Lauer:1998,Brown1998,Bender2005,Lauer2012}. In Chang's scenario, mass loss from giant stars and subsequent collisions between crossing gas streams lead to fresh star formation in a disk around the supermassive black hole every 500 Myr. The crossing orbits rely on a small precession rate ($\leq 3- 10$ \kmpersperpc) for the eccentric disk. 

Counter-rotation has been shown to be a critical dynamical driver of lopsided Kepler disks and has a significant impact on the precession rate \citep{Touma2002, Touma:2009, Sridhar2010, Touma2012, Kazandjian:2013, Touma2014}.
\citet{Kazandjian:2013} found that disks with a counter-rotating population with one-tenth the mass produces a disk that precesses at a near-constant rate of $\sim\!4$ \kmpersperpc. \citet{Sridhar2010} demonstrated no net rotation in the case of counter-rotating populations with identical radial profiles and masses,
which was then recovered by \cite{Touma2014} in their studies of maximum entropy equilibria for self-gravitating near-Keplerian disks.
The primary proposed mechanism in these studies for explaining the presence of a counter-rotating population of stars is accretion of a globular cluster that spirals into the galactic nucleus due to dynamical friction onto a pre-existing disk of stars around the black hole \citep{Sambhus2002}.
Alternatively, \citet{Hopkins10b} showed that eccentric nuclear disks may originate in gas-rich galaxy mergers, when gas that is funneled to the center of the potential fragments and forms stars on aligned eccentric orbits and can result in slow precession rates of 1-5 \kmpersperpc \citep[see also][]{Generozov:2022}. 
Both scenarios above involve external accretion into the nucleus to produce counter-rotation. 

\citet{Akiba:2021} showed that gravitational wave recoil kicks of merging black holes can reshape the surrounding stellar population into an apsidally-aligned eccentric disk. \citet{Akiba:2023} noted that stars at large distances from the black hole can reverse orbital orientation post-kick. They showed that counter-rotating populations will not have identical radial profiles even in regions of the disk that have equal masses in each. 
The previous study used massless stellar particles and did not explore the long-term evolution of such disks. 
In this paper, we evolve simple post-kick stellar systems in time including the self-gravity of the stars to explore the evolution of aligned eccentric disk of stars with varying fractions of counter-rotation.

In Section~\ref{sec:GW}, we describe how gravitational wave recoil kicks induce counter-rotation. In Section~\ref{sec:nbody} we describe the setup of our $N$-body simulations.
In Section~\ref{sec:results}, we present our results, specifically exploring disk alignment, eccentricity evolution, precession rate, and the tidal disruption event rate of eccentric nuclear disks produced after a black hole merger gravitational wave recoil kick. We show that significant counter-rotation induced by gravitational wave recoil kicks elevates all of these measures. We critically discuss our results in Section~\ref{sec:discussion}.

\begin{figure*}
 \includegraphics[width=\textwidth]{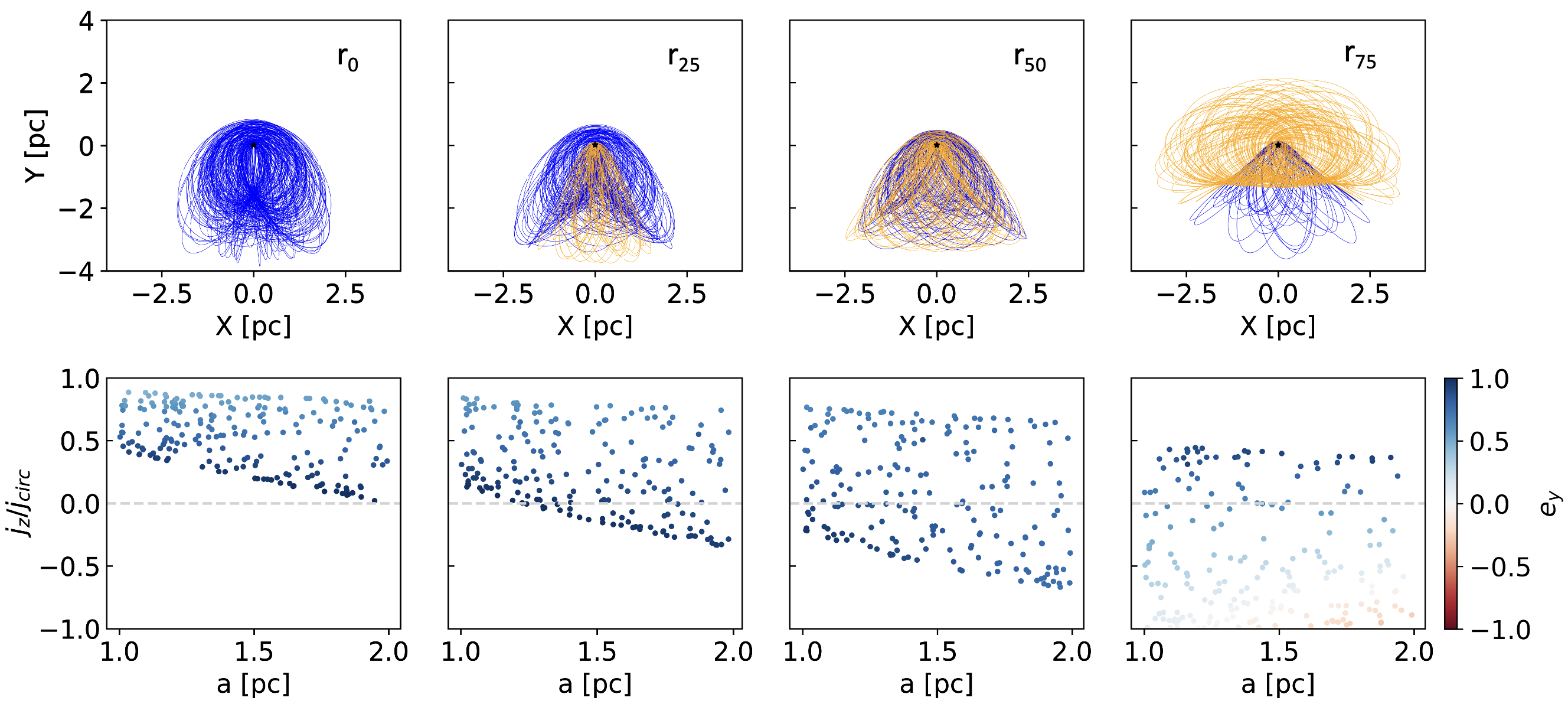}
 \caption{
 \textbf{Initial post-kick stellar orbits in four sets of $N$-body simulations ($r_{0}$, $r_{25}$, $r_{50}$, $r_{75}$).}
 \textit{Top:} Prograde (retrograde) orbits are in blue (orange). 
 \textit{Bottom:} z-component of the angular momentum of each star normalized by the circular angular momentum as a function of semi-major axis. The color bar shows the y-component of the eccentricity vector.}
     \label{fig:initial_orbits}
\end{figure*}

\section{Gravitational Wave Recoil Kicks}
\label{sec:GW}
High amplitude gravitational waves are emitted during the final stages of binary black hole inspiral and merger. Where there is asymmetry in the binary system (arising from unequal masses and/or spins), the emission is anisotropic, resulting in a gravitational wave recoil kick imparted on the remnant black hole \citep{Peres1962, Bekenstein1973, Wiseman1992, Campanelli2007, Herrmann2007}.

In \cite{Akiba:2021,Akiba:2023}, we investigated the effects of gravitational wave recoil kicks on orbiting stars. Here we expand on the case of an initially circular disk of stars and an in-plane recoil kick.
A circular stellar disk may form from a self-gravitating accretion disk \citep{Paczynski1978, Goodman2003}. The merger of two supermassive black holes leaves behind a flattened ellipsoid that is tangentially anisotropic in the inner regions \citep{Mastrobuono-Battisti2023}, for which a circular disk is a reasonable first-order approximation. The in-plane kick produces an apsidally-aligned eccentric disk, with the net eccentricity vector orthogonal to the kick direction in the $+y$ direction. 
The orbital motion of bound stars in the rest frame of the kicked black hole depends on which radial regime it initially lies in:
(1) where $\vc > \vk$, all stars become eccentric and remain on prograde orbits, 
(2) where $\vc < \vk$, orbits may become retrograde depending on their pre-kick anomaly, 
(3) where $2\vc < \vk$, retrograde orbits may become anti-aligned, i.e. their net eccentricity vectors point in the $-y$ direction. 
All bound orbits become retrograde in the frame of the moving black hole where $\vc \le \frac{1}{\sqrt{3}} \:\vk$. 
As $\vk$ increases, the semi-major axis range in which stars become retrograde post-kick moves inward, causing a larger total fraction of orbits to counter-rotate. 

\section{N-Body Simulations}
\label{sec:nbody}
\subsection{Methods}

We use the \texttt{IAS15} integrator in the $N$-body simulation package \texttt{REBOUND} \citep{Rein:2012, Rein:2015}. 
We use code units of $G=1$, $\Mbh =1$, and the inner edge of the disk $a_{\rm{in}} = 1$ such that the period of a circular orbit at the inner edge of the disk is $P(a_{\rm{in}}) = 2\pi$.
We initialize a circular stellar disk of stars in a semi-major axis range $a_{\rm{in}}=0.01$ and $a_{\rm{out}} =4$ with a surface density profile drawn from $\Sigma \sim a^{-1}$. 
Inclinations are normally distributed with a standard deviation of 0.05 radians.
We assume a recoil kick in the plane of the stellar disk as motivated by \citet{Bogdanovic2007}, suggesting that the alignment of orbits and black hole spins would skew the distribution toward more in-plane kicks. Since the pre-kick disk is axi-symmetric, we impart the kick in the $+x$ direction without loss of generality. For consistency of comparison across our four models, we vary the number of stars in the initial disk between $N= 10^3 - 10^5$ such that post-kick there remain $N\sim 200$ bound stars with post-kick orbits within a semi-major axis range $a \in [1,2]$ in code units with a disk mass of $M_{\rm{disk}} = 10^{-2} \Mbh$. We then select only the stars in this range to evolve in our simulations so that we can make comparisons between our models of analogous segments of the disk and limit the computational expense of our simulations.
As we are restricted in the number of N-body particles, each particle is over-massive compared to reality. As the ratio of the two-body relaxation timescale to the secular (orbit-averaged) timescale scales inversely with stellar mass, we expect our simulations to exhibit overly strong two-body scattering which perturbs particles off their orbits and weakens orbit-averaged torques. 
Post-Newtonian effects are excluded from these simulations since the precession timescale due to secular torques is several orders of magnitude shorter than the estimated general relativistic precession timescale \citep{Naoz:2016}. The dynamics and corresponding tidal disruption rates in eccentric disks are unaffected by the inclusion of post-Newtonian terms \citep{Wernke:2019}.

The magnitude of the gravitational wave recoil kick plays an important role in the post-kick retrograde fraction. 
We perform four different sets of simulations (five realizations each) varying the kick magnitude to produce post-kick stellar configurations with  0\% retrograde (i.e. all prograde), 25\% retrograde, 50\% retrograde, and 75\% retrograde orbits. Simulations are labelled $r_{0}$, $r_{25}$, $r_{50}$, $r_{75}$, respectively.  
The corresponding kick magnitudes in code units are $v_{\rm kick} = 0.5,  0.64,  0.82, 1.83$. 
The initial, post-kick, orbital configurations can be seen in the top panel of Fig. \ref{fig:initial_orbits}. The bottom panel shows the corresponding initial z-component of the orbital angular momentum normalized by the circular angular momentum, $\jzjc$, for each star. $\jzjc > 0$ are prograde orbits and $\jzjc < 0$ are retrograde. 
The $r_{50}$ simulations have a very small positive net angular momentum, $\jzjc > 0$. 
Unless otherwise specified, we scale the mass of the black hole to $\Mbh = 10^8 M_{\odot}$, and the inner edge of the disk to  $1 \rm{pc}$, when calculating physical timescales, loosely modeled on the M31 nucleus. Kick magnitudes in this case map to 
$v_{\rm kick} = 328, 420, 538, 1200 \ \rm{km} \ \rm{s}^{-1}$. Typical kick velocities due to anisotropic gravitational waves are $\mathcal{O}(100 \ \rm{km} \ \rm{s}^{-1})$, but can reach several thousand kilometres per second \citep{Favata2004, Blanchet2005, Damour2006, Campanelli2007}.

\section{Results}
\label{sec:results}

\subsection{Oscillations in Stellar Angular Momentum}
\label{sec:j}

Orbit orientation is given by the eccentricity vector 
\begin{equation}
\vec{e} = \frac{\vec{v} \times \vec{j}}{G \Mbh} - \frac{\vec{r}}{r}
\label{eq:e}
\end{equation}
where $\vec{r}$ is the radius vector, $\vec{v}$ is the velocity vector, $|\vec{j}| = |\vec{r} \times \vec{v}| = 
(G \Mbh a(1-e^2))^{1/2}$  is the specific angular momentum vector, $a$ is the semi-major axis, $e$ is the eccentricity, and $\Mbh$ is the mass of the central black hole.  
The precession of an orbit is defined by the rate of change of the eccentricity vector
\begin{equation}
  e' = \frac{\vec{f} \times \vec{j} }{G \Mbh} + \frac{\vec{v} \times \vec{\tau}}{G \Mbh   }
\end{equation}
where $\vec{f}$ is the (specific) non-Keplerian gravitational force, $\vec{\tau} = dj/dt$ is the specific orbit-averaged torque. 

\begin{figure}
    \centering
        \includegraphics[scale=.5]{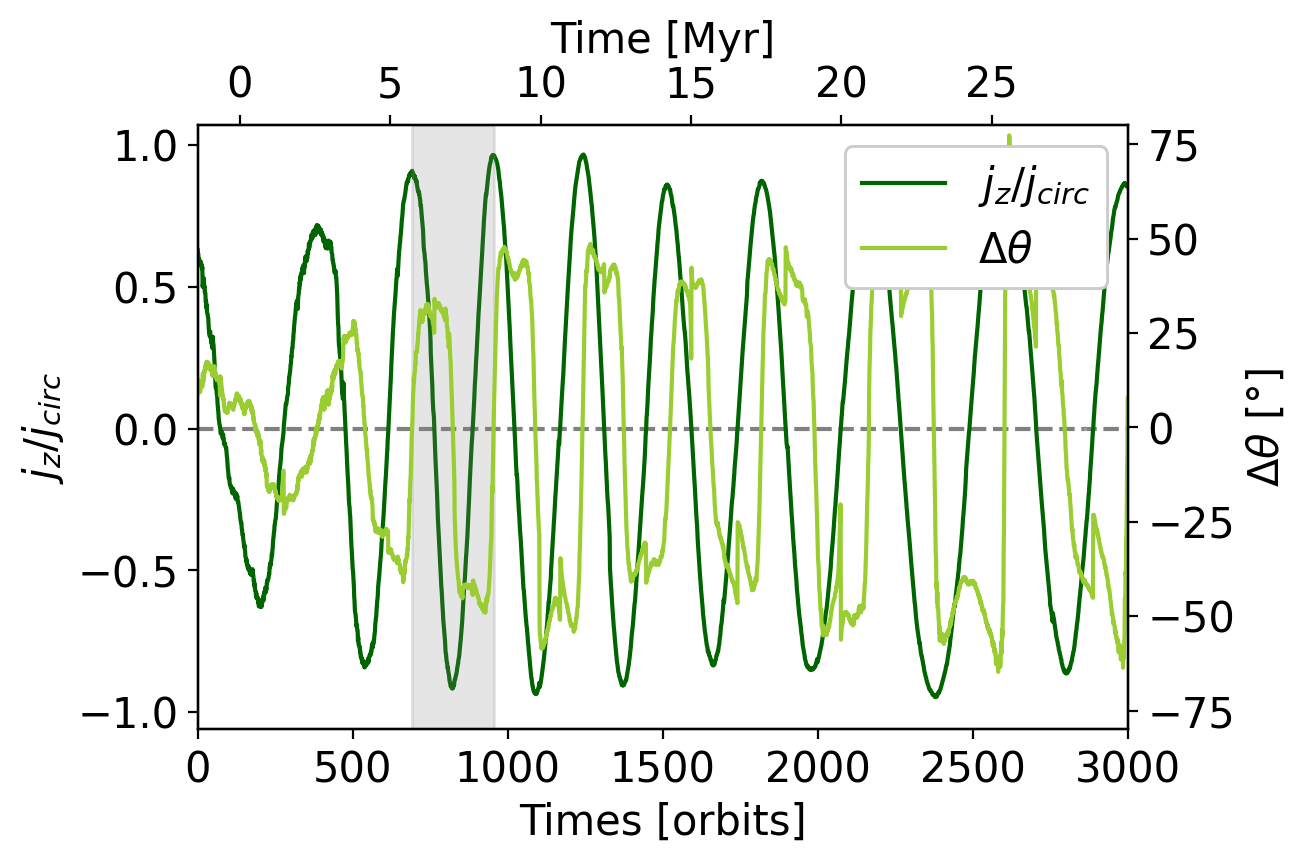}
        \includegraphics[scale=.5]{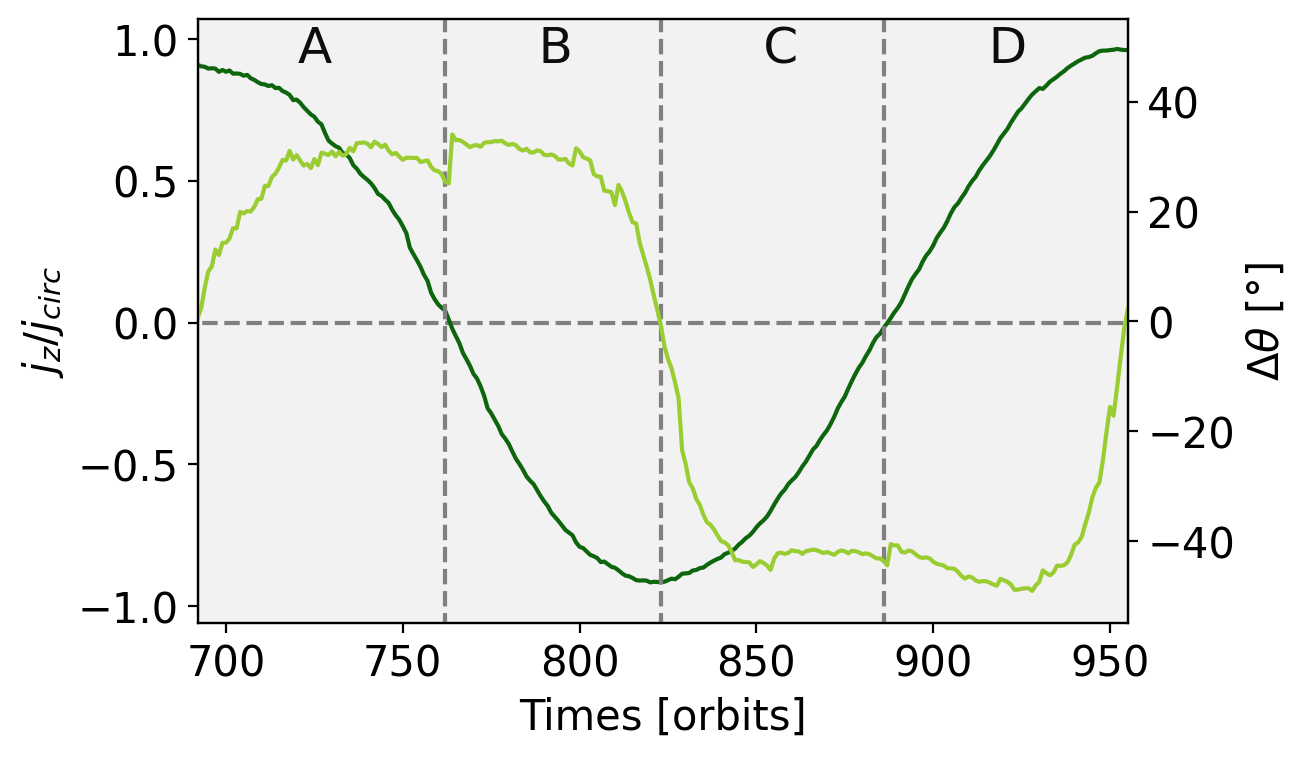}
    \caption{
    \textbf{Flips in orbit orientation} 
    \textit{Top:} Time evolution of the $z$-component of the angular momentum, normalized to the circular angular momentum, for a sample star from the $r_{50}$ model, and $\Delta \theta$ defined to be the azimuthal angle between the star's eccentricity vector and the average eccentricity vector of the disk. 
    \textit{Bottom:} A zoom in of the shaded region of the top figure to illustrate the four different possibilities of the sample star's configuration relative to the bulk of the disk.}
    \label{fig:jz}
\end{figure}

 When a given stellar eccentricity vector is misaligned with the average eccentricity vector of the massive disk, it will experience a strong gravitational torque. If this torque is positive (relative to the orbit's angular momentum) it will serve to circularize the orbit, whereas if it is negative it will serve to increase the orbit's eccentricity. In Fig. \ref{fig:jz} we show the time evolution of the $z$-component of the angular momentum normalized by the circular angular momentum, $\jzjc$, of a single star in the $r_{50}$ simulation. We also show $\Delta \theta$, the azimuthal angle between the star's eccentricity vector and the average eccentricity vector of the disk. The bottom panel shows a zoom-in of the shaded region in the top panel chosen for illustrative purposes to demonstrate the four possible states of the stellar orbit relative to the disk:
\begin{enumerate}[(A)]
\item When $\jzjc > 0$ (i.e. the star's angular momentum is prograde with respect to the disk) and $\Delta \theta >0$ (i.e. the orbit has precessed ahead of the disk) the massive disk induces a negative torque on the star's orbit. This decreases the orbital angular momentum until $\jzjc $ passes through zero and the orbit flips to a retrograde orientation. 
\item When $\jzjc < 0$ (i.e. the star's orbit is retrograde) it's precession is in the opposite direction, which acts to decrease $\Delta \theta$. The torque from the disk is now positive with respect to the stellar angular momentum, thereby decreasing its orbital eccentricity. The stellar orbit precesses past the bulk of the disk resulting in $\Delta \theta < 0$. 
\item When $\jzjc <0$ and $\Delta \theta <0$, the disk induces a negative torque on the star's angular momentum. This increases its orbital eccentricity and decreases $|\jzjc |$ until it flips orientation to become prograde. This is the inverse of (A). 
\item When $\jzjc >0$, its precession is in the prograde direction, bringing it into alignment with the bulk of the disk and decreasing $|\Delta \theta|$. The bulk of the disk induces a positive torque on the star's orbit. This decreases the orbital eccentricity and increases $\jzjc $. This is the inverse of (B). 
\end{enumerate}

Orbit-averaged torques from the massive disk cause stellar orbits to oscillate continuously in angular momentum, flipping between prograde and retrograde (see top panel of Fig. \ref{fig:jz}).
These oscillations contain the disk orbits in tight apsidal alignment (Section \ref{sec:eccentricity}) and result in extreme tidal disruption event rates (Section \ref{sec:TDE}). This torquing mechanism is strongest in our $r_{50}$ simulations in which the initial post-kick disks are apsidally aligned and the equal-mass prograde and retrograde populations begin to precess in opposite directions. Strong inter-orbit torques between the two oppositely-precessing populations cause both populations to flip orientation and precess back towards each other. These dynamics repeat continuously. 

\subsection{Disk Alignment and Eccentricity Evolution}
\label{sec:eccentricity}

\begin{figure}
    \centering
    \includegraphics[width=\columnwidth]{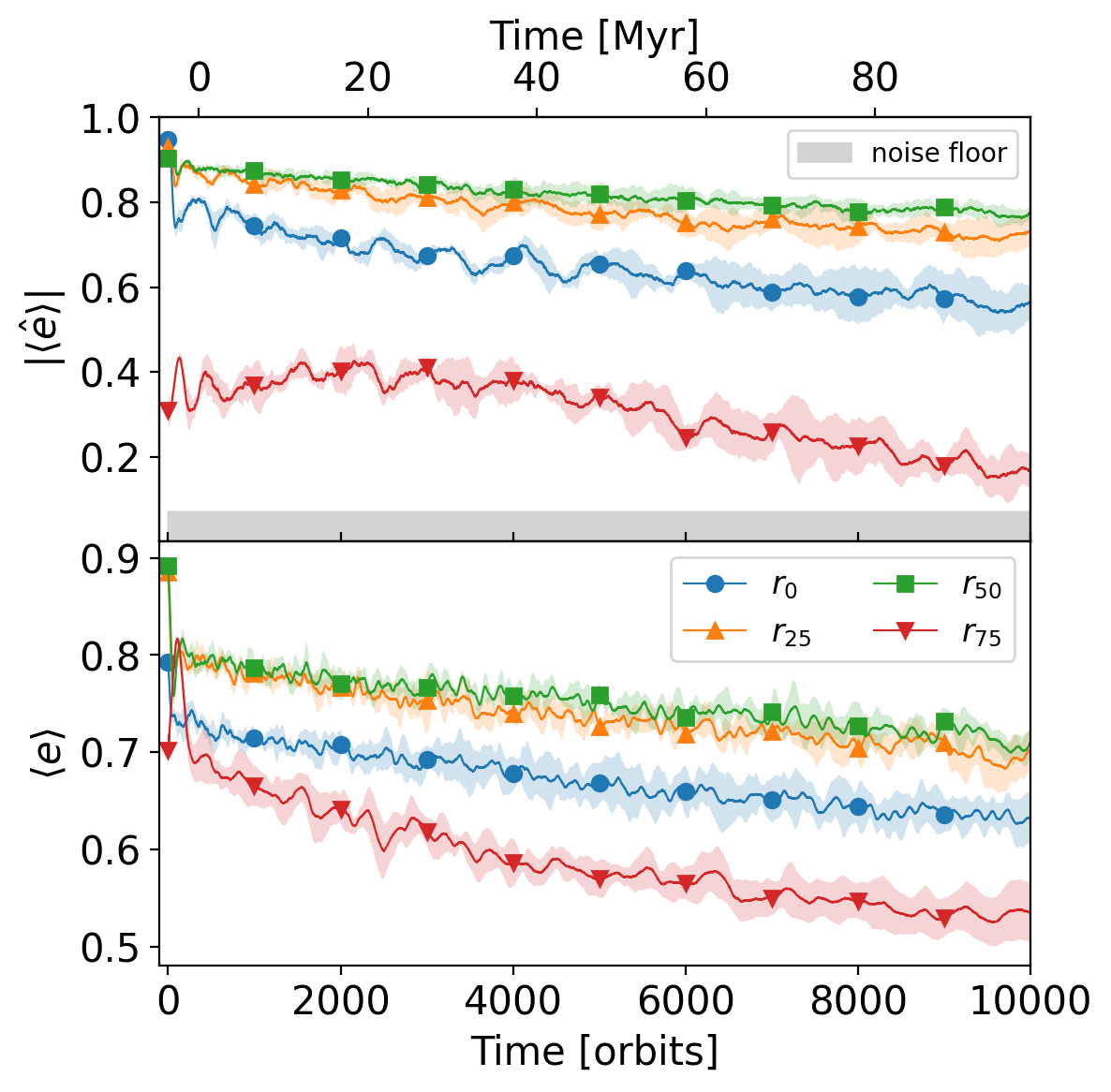}
    \caption{\textbf{Disk Alignment and Average Eccentricity Evolution.} 
    \textit{Top:} Time evolution of the average unit eccentricity vector of stars in the disk. Different colored lines correspond to models of varying retrograde fraction. The noise floor, calculated as $1/\sqrt{N_{\rm bound}(t=0)}$ is shown with a grey shaded region.
    \textit{Bottom:} Time evolution of the average eccentricity of all the stars in the disk.}
    \label{fig:ecc}
\end{figure}

We quantify the apsidal alignment of the disk using the average unit eccentricity vector 
\begin{equation}
    \langle \hat{e} \rangle = \frac{\sum_{i=1}^{N_{\rm{bound}}} \hat{e}_i}{N_{\rm{bound}}}
\end{equation}
where $ \langle \hat{e} \rangle = 1 $ indicates maximum apsidal alignment. The top panel of Fig. \ref{fig:ecc} shows the time evolution of the unit eccentricity vector for each of our simulations, showing the average (solid line) and standard deviation (shaded region) of the five realizations of each simulation set. Alignment is greatest for our $r_{50}$ and $r_{25}$ models, and lowest for our $r_{75}$. 
The $r_{75}$ model has the lowest alignment because it starts with a misalignment between the prograde and retrograde orbits (see Fig. \ref{fig:initial_orbits}). 

We show the average orbital eccentricity as a function of time in the bottom panel of Fig. \ref{fig:ecc}. All our simulations start with a large average eccentricity due to the impact of the recoil kick (pre-kick eccentricities are zero). 
Eccentricities remain very high when there is a significant retrograde fraction within the disk, with the largest values again in the $r_{50}$ and $r_{25}$ models. This is due to the strong negative torques (relative to the orbital angular momentum) between the two massive counter-rotating populations which drive a rapid decrease in the absolute value of angular momentum of each population, driving orbits to high eccentricities. While the $r_{50}$ and $r_{25}$ models have very similar average eccentricities, the distribution of those eccentricities is different. In the $r_{50}$ models the prograde and retrograde populations have the same average eccentricity, while in the $r_{25}$ models the retrograde population has a higher average eccentricity. This is initially due to the orbital distribution caused by the recoil kick: in the $r_{25}$ model the only orbits that are flipped onto retrograde orbits are also highly eccentric, whereas in the $r_{50}$ more orbits are flipped onto retrograde orbits and have a wider range of eccentricities due to the fact that $v_{\rm kick}$ is greater and therefore there are more orbits for which $v_{\rm{circ}} \ll v_{\rm kick}$ so there can be retrograde orbits with lower eccentricities (see Fig. \ref{fig:initial_orbits}). 

We note that the retrograde fraction at the end of our simulations is ($r_{0},r_{25},r_{50},r_{75}$ = {0.17, 0.34, 0.47, 0.82}). The $r_{0}$ and $r_{25}$ simulations increase in retrograde fraction due to the fact that the inter-orbit torques cause some orbits to flip orientation, and this initially occurs primarily to the prograde population as it makes up the majority (or in the case of $r_0$ the entirety) of the disk \citep[see also][]{Rantala:2023}. The retrograde fraction in the $r_{25}$ simulation increases while it slightly decreases for the  $r_{50}$ simulation, accounting for their similar evolution. 

The $r_{75}$ simulation displays a qualitatively different behavior. Due to the larger kick magnitude, the $r_{75}$ starts with eccentricity vectors pointed in many different directions (see Fig. \ref{fig:initial_orbits}), and therefore there is initially a much lower amount of apsidal alignment in the disk (see Fig. \ref{fig:ecc}). Inter-orbit torques are correspondingly weaker and less coherent. The prograde orbits are initially much more eccentric on average than the retrograde orbits (see Fig. \ref{fig:initial_orbits}), leading more of them to flip due to the inter-orbit torques and cause an increase in the retrograde fraction. 
Due to the initial apsidal spread of the orbits, the torques fail to align the orbits, and the average unit eccentricity vector, $\langle \hat{e} \rangle$, drops with time due to the differential precession.

\subsection{Precession Rate}
\label{sec:precession}

\begin{figure}
    \centering
    \includegraphics[width=0.45\textwidth]{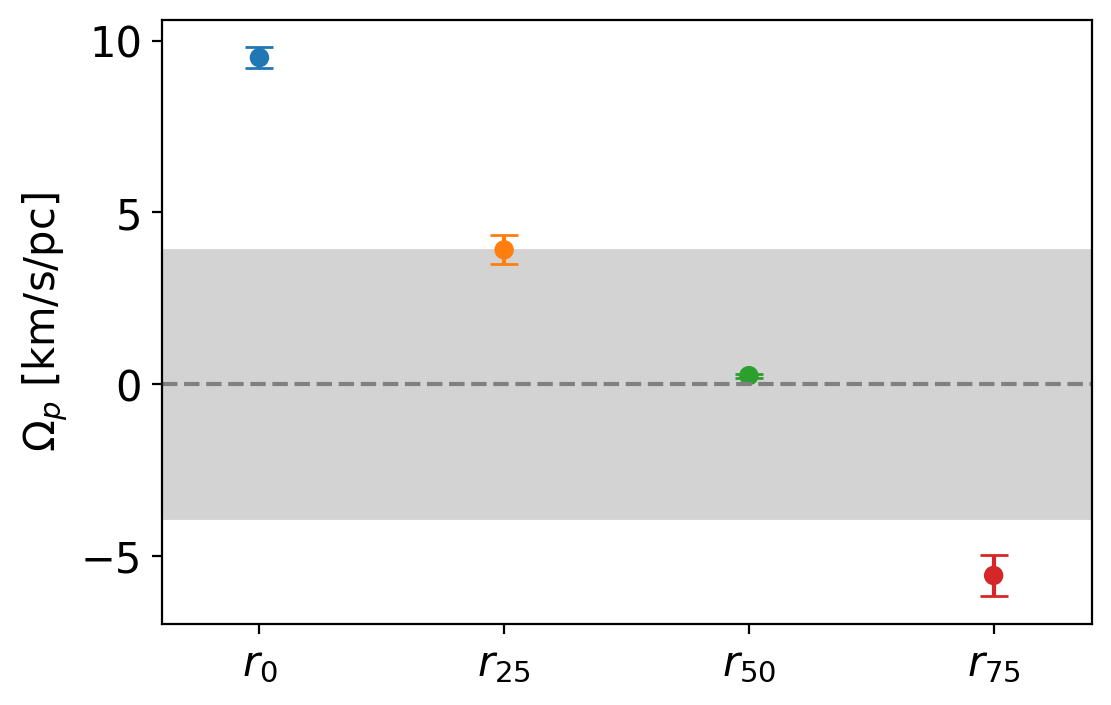}
    \caption{
    \textbf{Precession rates as a function of retrograde fraction.} 
    The dashed line and shaded region indicates the range of precession rates that are consistent with the M31 eccentric nuclear disk \citep{Lockhart:2018}.}
    \label{fig:precession}
\end{figure}

We define the precession rate of the disk, $\Omega_P$, as the time it takes the average eccentricity vector of the disk to complete a full $2\pi$ rotation. 
We take the average and standard deviation over the five realizations of each simulation set.
To compare with the M31 nucleus, we augment the disk mass by a factor of ten. This augmentation is performed in post-processing due to computational limitations. The M31 eccentric disk is approximately an order of magnitude larger in both radial extent and mass, so this extrapolation is necessary to make meaningful comparisons with observed precession rates. The precession rates for each of our simulations can be seen in Fig. \ref{fig:precession}. The range of values for the precession rate of the eccentric nuclear disk in M31 is shown as a grey shaded region \citep{Lockhart:2018}. 

In our $r_{50}$ models, the precession rate of the aligned disk is nearly zero, in excellent agreement with \citet{Lockhart:2018}. The closer the retrograde fraction is to 50\%, the slower the precession rate. 
In simulations with more of one orientation there is a net precession of the aligned disk in the direction dictated by the majority population. 
The precession rates for each of our simulations can be seen in Fig. \ref{fig:precession}.
Our work provides a new mechanism to explain the slow precession rate in the M31 eccentric nuclear disk: a significant retrograde fraction caused by a recoil kick serves to keep the disk static, as well as maximally aligned.

\subsection{Tidal Disruption Events}
\label{sec:TDE}

Tidal disruption events (TDEs) occur when a star passes within the tidal disruption radius of the black hole given by $R_T = R_* (\Mbh/M_*)^{1/3}$ \citep{Rees:1988}. Since the closest approach of a star is given by the periapsis distance $R_p = a(1-e)$, large eccentricities result in significantly closer encounters with the black hole and increase the likelihood of disruption. TDE rates have been shown to be greatly enhanced in eccentric nuclear disks due to gravitational torques from the aligned disk driving oscillations in eccentricity \citep{Madigan:2018, Wernke:2019, Wernke:2021}. 

Here we explore the TDE rate in our four simulation sets. 
The supermassive black hole in M31 is too massive to produce observable TDEs, as its Schwarzchild radius is larger than the tidal disruption radius. We therefore scale our black hole mass to $10^7 M_{\odot}$,  $a_{\rm{in}} = 0.1 \ \rm{pc}$, and we scale the stellar masses and radii to that of the sun in order to calculate the tidal disruption radius.

\begin{figure}
    \centering
            \includegraphics[scale=.55]{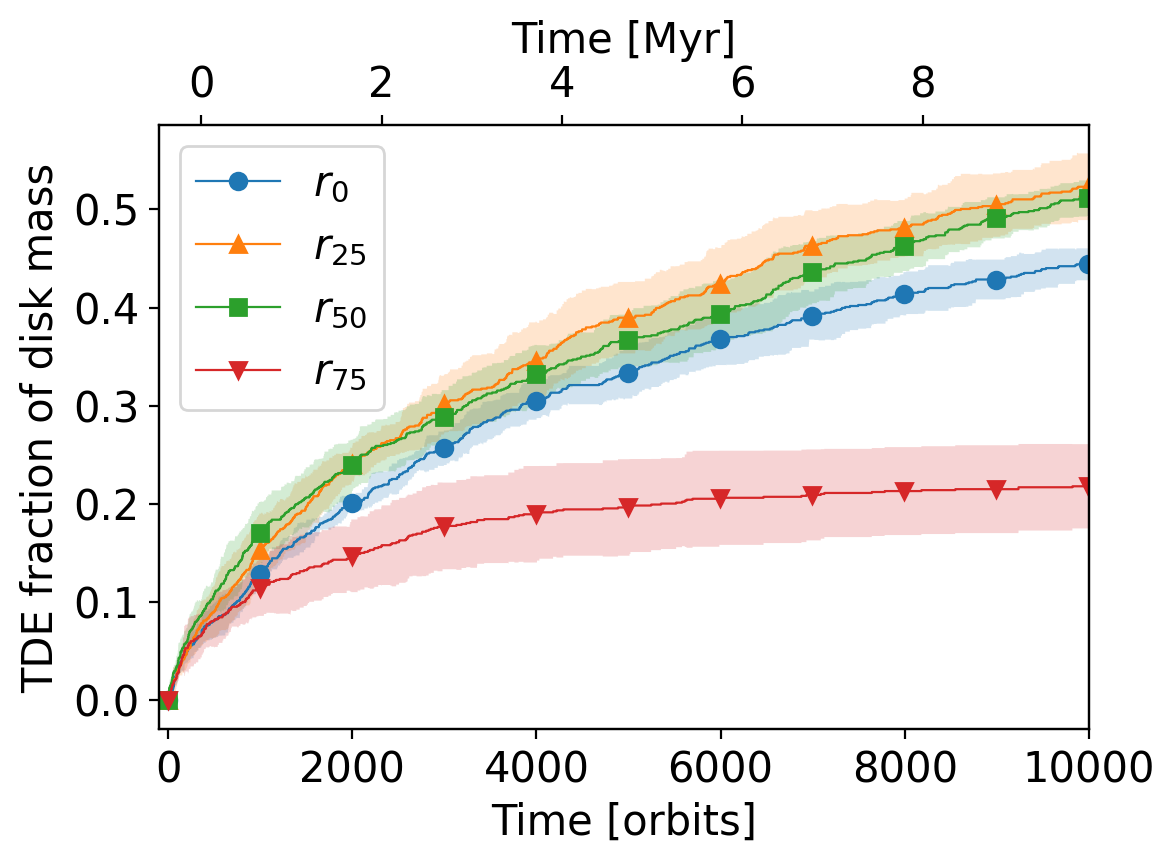}
    \caption{
    \textbf{Fraction of the disk that has experienced a tidal disruption event as a function of time.}}
    \label{fig:TDE}
\end{figure}

We show our results in Fig. \ref{fig:TDE}, plotting the fraction of stars in the disk that undergo a TDE as a function of time. We find a high TDE rate in all of our simulations that maintain high apsidal-alignment.
This is due to both the high average eccentricity in these simulations and the periodic gravitational torques which cause angular momentum values to pass through zero as orbit orientations flip. 
We see significantly lower TDE rates in the $r_{75}$ simulation as this disk does not maintain high apsidal-alignment or large eccentricities. 
Sufficiently high rates of TDEs will not appear as distinct events but rather have disrupting stars interacting with pre-existing TDE accretion disks. The build-up of numerous such disks may mimic the appearance of an AGN, providing an electromagnetic counterpart to the gravitational wave signal of the black hole merger. 
We note that in these simple numerical experiments, we do not remove stars from our simulations after they experience a TDE, but simply count the first time that a star passes within the black hole tidal disruption radius. 
In reality, TDEs will remove mass but little angular momentum from the stellar disk, causing it to gain angular momentum per unit mass. Unless the disk mass can be replenished from the surrounding stellar potential, this will have two effects. One is to slowly circularize the eccentric nuclear disk and, two, to decrease the TDE rate over time.

\section{Discussion}
\label{sec:discussion}

We perform $N$-body simulations to investigate the long-term evolution of eccentric stellar nuclear disks around supermassive black holes formed due to gravitational wave recoil kicks.
We simulate disks with both prograde and retrograde orbits, varying the magnitude of the recoil kick and thus the fraction of retrograde orbits within the disk. We describe the mechanism by which retrograde orbits help to maintain the disk's apsidal alignment through orbit-averaged torques between oppositely precessing prograde and retrograde populations.  
Our main findings are as follows:
\begin{enumerate}
    \item Gravitational wave recoil kicks can produce any distribution of retrograde fraction in the orbits of a disk given different kick magnitudes, or initial stellar configurations. This is a new mechanism by which counter-rotation may be introduced into a galactic nucleus. 

    \item Large retrograde fractions are induced by large (relative to the pre-kick stellar orbital velocities) gravitational wave recoil kick velocities. 

    \item Aligned disks with a significant fraction of retrograde orbits are the most dynamically stable (Fig.~\ref{fig:ecc}). 

    \item Aligned disks with a significant fraction of retrograde orbits are the most eccentric (Fig.~\ref{fig:ecc}). 

    \item Aligned disks with an equal fraction of prograde and retrograde orbits have near zero precession rates (Fig~\ref{fig:precession}), consistent with that of M31's eccentric nuclear stellar disk \citep{Lockhart:2018}

    \item Tidal disruption rates are greatly enhanced in eccentric nuclear disks with high apsidal-alignment (Fig.~\ref{fig:TDE}). 

\end{enumerate}
Gravitational wave recoil kicks spontaneously generate counter-rotating orbits in nuclear star clusters. This is an interesting new explanation for nuclei with significant counter-rotation \citep[e.g.,][]{Gultekin2014}. If this is the reason behind the near-zero precession rate of the M31 nuclear disk, our work also predicts that this disk contains a significant fraction of stars and stellar remnants on retrograde orbits. \cite{Sambhus2002} find that a small fraction of retrograde orbits in their disks improves the fits to the observed kinematics and photometry of the M31 nucleus. 
We do not make a similar comparison here as we are simulating a narrow range in semi-major axes. In future, we will explore the impact of gravitational wave recoil kicks on the dynamics of realistic pre-kick stellar distributions scoured by merging supermassive black holes \citep{Mastrobuono-Battisti2023} across a wider range of semi-major axes.

\section*{Acknowledgements}

We gratefully acknowledge support from the David and Lucile Packard Foundation and NSF Graduate Research Fellowship grant DGE-1746060. 

\section*{Data Availability}

The data underlying this article will be shared on reasonable request to the corresponding author.

\bibliographystyle{mnras}
\bibliography{Bibliography}

\end{document}